	\documentclass[prl,amssymb,amsmath,preprintnumbers,showpacs,twocolumn]{revtex4}

	\usepackage{graphicx}
	\usepackage{color}
	\usepackage{verbatim}
	\usepackage{subfigure}
	\include{subcaption}
	\usepackage{amssymb}

\begin{document}

	\title{Maxwell's Refrigerator: An Exactly Solvable Model}
	\author{Dibyendu Mandal$^1$, H. T. Quan$^{2,3}$ and Christopher Jarzynski$^{2,4}$}
	\affiliation{ 
	$^1$Department of Physics, University of Maryland, College Park, Maryland 20742, U.S.A.\\
	$^2$Department of Chemistry and Biochemistry, University of Maryland, College Park, Maryland 20742, U.S.A.\\
	$^3$School of Physics, Peking University, Beijing 100871, China.\\
	$^4$Institute for Physical Science and Technology, University of Maryland, College Park, Maryland 20742, U.S.A.}

	\begin{abstract}
	
We describe a simple and solvable model of a device that -- like the ``neat-fingered being'' in Maxwell's famous thought experiment -- transfers energy from a cold system to a hot system by rectifying thermal fluctuations.
In order to accomplish this task, our device requires a memory register to which it can write information: the increase in the Shannon entropy of the memory  compensates the decrease in the thermodynamic entropy arising from the flow of heat against a thermal gradient.
We construct the nonequilibrium phase diagram for this device, and find that it can alternatively act as an eraser of information.
We discuss our model in the context of the second law of thermodynamics.

	\end{abstract}

	\maketitle

In a thought experiment highlighting the statistical nature of the second law of thermodynamics, Maxwell imagined a tiny creature acting as a gatekeeper between two  chambers filled with gases at different temperatures.
By preferentially allowing fast-moving molecules to pass from the cold to the hot chamber, and slow ones to pass in the other direction, this creature achieves refrigeration without expending energy.
As Maxwell put it: 
``the hot system has got hotter and the cold colder and yet no work has been done, only the intelligence of a very observant and neat-fingered being has been employed''~\cite{Leff2003}.

\begin{figure}
	\includegraphics[trim = 0in 0in 0in 0in, clip = true, scale = 0.3]{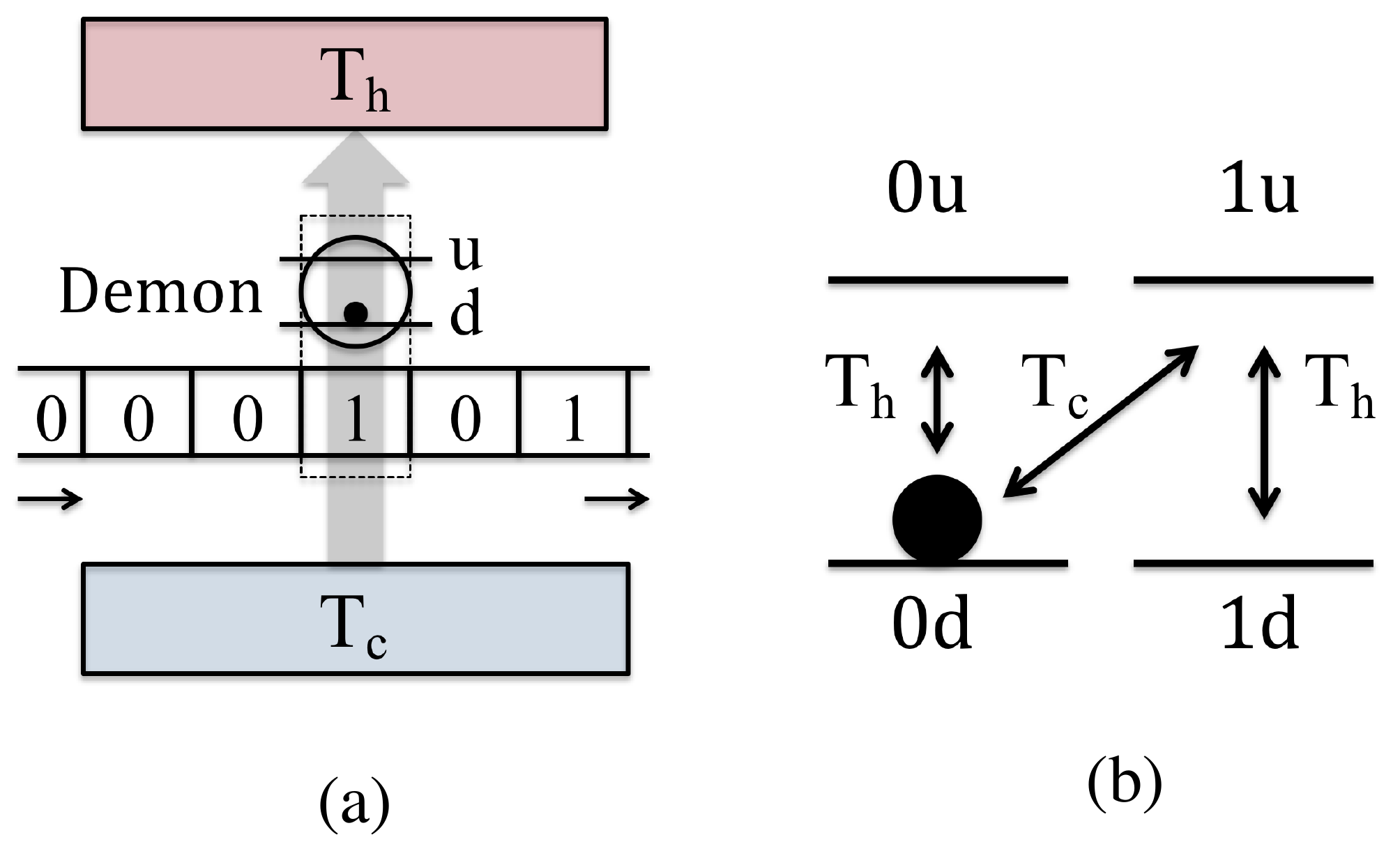}
	\caption{
	(a) The device, or ``demon'', interacts with a sequence of bits, one at a time, while exchanging energy with two thermal reservoirs.
	(b) The demon makes intrinsic transitions mediated by the hot reservoir (vertical arrows), and the demon and nearest bit make cooperative transitions $0d \, \leftrightarrow \, 1u$ mediated by the cold reservoir (diagonal arrows).}
	\label{fig:Setup}
	\end{figure}

In this Letter we propose a simple, solvable model of a physical device that accomplishes the same result as Maxwell's intelligent and observant creature: it creates a flow of energy against a thermal gradient, without the input of external work.
Our device is a classical two-state system that interacts with a pair of thermal reservoirs and a memory register, which we model as a stream of bits (Fig.~\ref{fig:Setup}(a)).
The dynamics consist of stochastic transitions, by means of which the device exchanges energy with the reservoirs and modifies the states of the bits.
For appropriate values of the model parameters, these dynamics produce a steady state in which there is a continual flow of energy from the cold reservoir to the hot reservoir, and a record of the system's microscopic evolution is continually written to the stream of bits.
Our device is fully autonomous, requiring no intervention by an external agent.
Its ability to control the flow of energy between the reservoirs emerges entirely from the microscopic equations of motion.

The term ``Maxwell's demon'' has come to refer not only to the original setting described by Maxwell, but more generally to any situation in which a rectification of microscopic fluctuations produces a decrease of thermodynamic entropy~\cite{Maxwell1871,Szilard1929}.
A consensus has emerged that a physical device could achieve such a result, without violating the second law,
if it were simultaneously to write information to a memory register~\cite{Landauer1961,Penrose1970,Bennett1982,Bennett1985,Maroney2009}.
In this view, the act of writing increases the information entropy of the memory register, thereby compensating the decrease of thermodynamic entropy produced by the device.
If the information is later erased from the memory register, then by Landauer's principle~\cite{Landauer1961,Berut2012} there must be an increase in thermodynamic entropy elsewhere.
This tidy accounting places the Shannon entropy of a sequence of bits on the same thermodynamic footing as the Clausius entropy, defined in terms of heat and temperature.
As long as the sum of these entropies never decreases, the second law remains satisfied.
See, however, Refs.~\cite{Earman1998, Earman1999, Hemmo2010,Norton2011} for dissenting perspectives, which suggest that this consensus is at best an appealing narrative based on the presupposition of the second law, rather than an independent explanation.

Maxwell's demon has recently enjoyed increased attention in a broad range of settings, including artificial molecular machines~\cite{Kay2007}, single photon cooling of atoms~\cite{Raizen2011}, biomolecular signal transduction~\cite{Tu2008}, quantum information theory~\cite{delRio2011} and the feedback control of microscopic fluctuations~\cite{Kim2007, Sagawa2008, Sagawa2009, Jacobs2009,Cao2009, Sagawa2010, Ponmurugan2010, Horowitz2010,Toyabe2010, Abreu2011, Vaikuntanathan2011, Dong2011,Sagawa2012,Jacobs2012,Abreu2012,Sagawa2012a}.
Maxwell's nineteenth-century thought experiment has become a touchstone for discussing the thermodynamic implications of information processing by physical systems~\cite{Zurek1989,Bub2001,Maruyama2009,Hosoya2011}.
While the consensus described above has identified and clarified these implications, far less effort has been devoted to uncovering precisely {\it how} a physical device, acting on its own, might accomplish the same result as Maxwell's hypothetical being~\cite{Quan2006,Bier2012,Mandal2012,Strasberg2013,Horowitz2012,Barato2013}.
To the best of our knowledge, the autonomous model we introduce below is the first to generate a flow of energy against a thermal gradient, effectively acting as a refrigerator without a power supply -- just as in the setup  considered by Maxwell, but with the intelligent creature replaced by a dumb device.
This contrasts with an earlier model of a device that acts as an {\it engine}, supplying work by extracting heat from a single thermal reservoir~\cite{Mandal2012}. 
Our autonomous framework also differs from that of Refs.~\cite{Kim2007, Sagawa2008, Sagawa2009, Jacobs2009,Cao2009, Sagawa2010, Ponmurugan2010, Horowitz2010,Toyabe2010, Abreu2011, Vaikuntanathan2011, Dong2011,Sagawa2012,Jacobs2012,Abreu2012,Sagawa2012a} (including the experimental realization reported in Ref.~\cite{Toyabe2010}), in which external intervention in the form of measurement and feedback is a key element.

\begin{figure}
	\includegraphics[scale = 0.35]{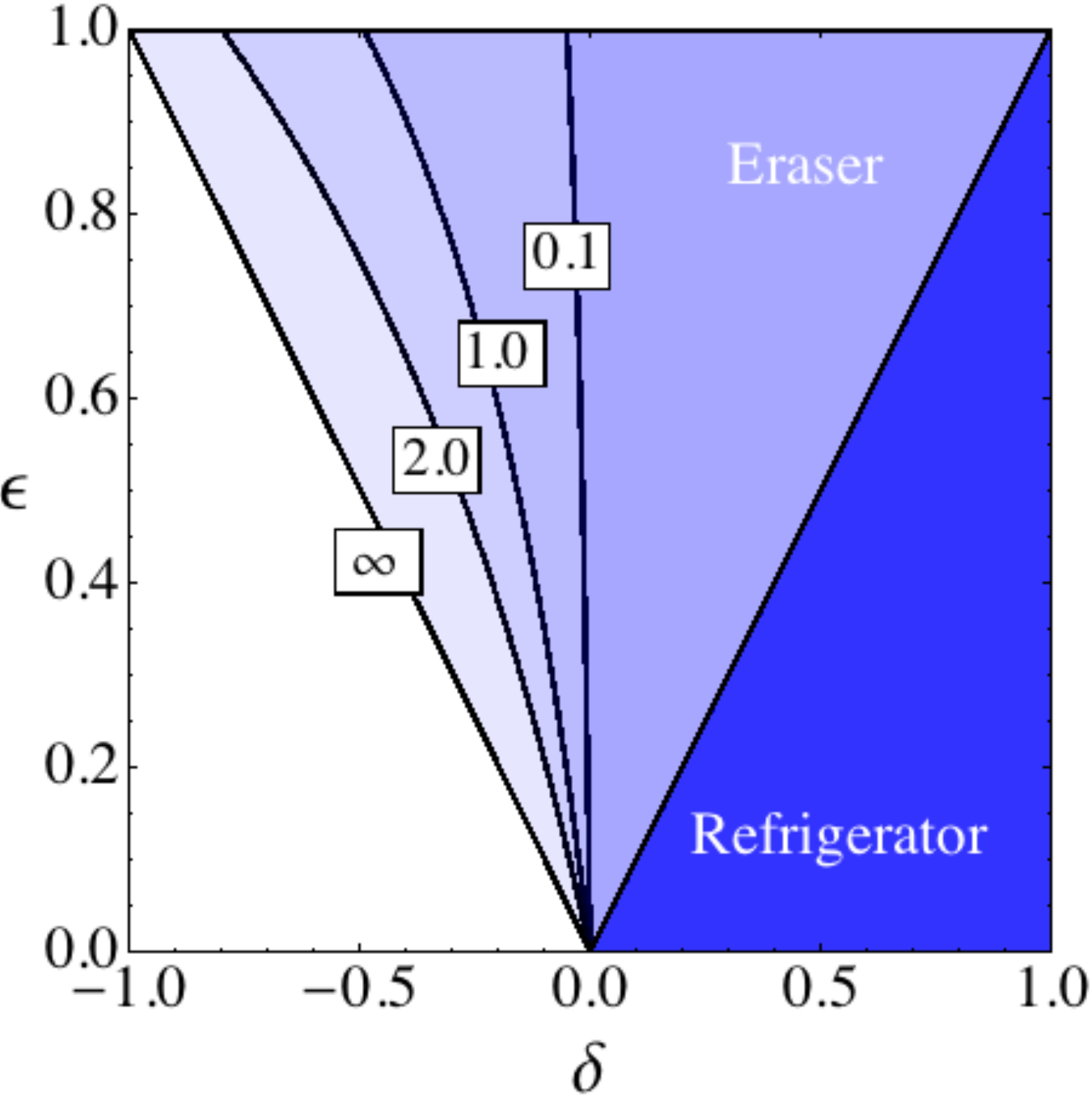}
	\caption{Phase diagram of our model at fixed $\gamma=1$ and $\omega=1/2$.
	The parameter $\delta$ specifies the incoming bit statistics, and $\epsilon$ is a rescaled temperature difference (Eq.~\ref{eq:Epsilon}).
	In the most darkly shaded region the demon acts as a refrigerator ($Q_{c\rightarrow h}>0$), while in the lightly shaded regions it acts as an eraser ($\Delta S_B<0$).
	The left boundary of the eraser region is shown for $\tau = 0.1$, $1.0$, $2.0$ and $\infty$.
	In the blank region at the lower left, our model exhibits neither behavior (see text).}
	\label{fig:Phase}
	\end{figure}

In what follows we describe our model and analyze its dynamics.
We obtain a nonequilibrium phase diagram for the steady state behavior (Fig.~\ref{fig:Phase}), which reveals that our device can act either as a refrigerator, transferring energy from a cold to a hot reservoir, or as an eraser, decreasing the information content of the memory register.
Finally, we briefly discuss our model in the context of the second law of thermodynamics.

Our model consists of four components, sketched in Fig.~\ref{fig:Setup}(a): a memory register, two thermal reservoirs at temperatures $T_c$ and $T_h> T_c$, and a device that plays the role of Maxwell's demon.
The memory register is a sequence of bits (two-state systems) spaced at equal intervals along a tape that slides frictionlessly past the demon.
The demon interacts with the nearest bit and with the reservoirs, as we describe in detail in the following paragraphs.

The demon itself is a two-state system, with states $u$ and $d$ characterized by an energy difference $\Delta E = E_u-E_d>0$.
It can make random transitions between these two states by exchanging energy with the hot reservoir, as illustrated by the vertical arrows in Fig.~\ref{fig:Setup}(b).
We will refer to these as {\it intrinsic} transitions, to emphasize that they involve the demon but not the bits.
The corresponding transition rates satisfy the requirement of detailed balance~\cite{vanKampen2007},
 	\begin{equation}
	\label{eq:Boltzmann_Th}
	\frac{R_{d \rightarrow u}}{R_{d \leftarrow u}} = e^{- \beta_h \Delta E},
	\end{equation}
where $\beta_h = 1/ k T_h$ and $k$ is Boltzmann's constant. 
We parametrize these rates as
	\begin{equation}
	\label{eq:sigma_gamma}
	R_{d \rightarrow u} = \gamma (1 - \sigma), \quad R_{d \leftarrow u} = \gamma (1 + \sigma), \quad \sigma = \tanh{\frac{\beta_h \Delta E}{2}}
	\end{equation}
where $\gamma>0$ sets a characteristic rate for these transitions, and $0<\sigma<1$.

Each bit has two states, 0 and 1, with equal energies.
We assume there are no intrinsic transitions between these two states.
That is, the state of the bit can change only via interaction with the demon, as we now discuss.

At any instant in time, the demon interacts only with nearest bit.
As a result, it interacts sequentially with the bits as they pass by.
The duration of interaction with each bit is $\tau = l/v$, where $l$ is the spacing between bits and $v$ is the constant speed of the tape.
During one such {\it interaction interval}, the demon and the nearest bit can make {\it cooperative} transitions: if the bit is in state 0 and the demon is in state $d$, then they can simultaneously flip to states 1 and $u$, and vice-versa (Fig.~\ref{fig:Setup}(b), diagonal arrows).
We will use the notation $0d \leftrightarrow 1u$ to denote these transitions, which are accompanied by an exchange of energy with the cold reservoir.
The corresponding transition rates again satisfy detailed balance,
	$R_{0d \rightarrow 1u}/R_{0d \leftarrow 1u} = e^{-\beta_c \Delta E}$,
where $\beta_c = 1/ k T_c$, and we will parametrize them as follows~\footnote{Note the lack of a rate parameter analogous to $\gamma$ in Eq.~\ref{eq:sigma_gamma}.  For the cooperative transition rates, we set this parameter to unity by appropriately choosing the unit of time.}
	\begin{equation}
	\label{eq:omega}
	R_{0d \rightarrow 1u} = 1 - \omega , \quad R_{0d \leftarrow 1u} = 1 + \omega , \quad \omega = \tanh{\frac{\beta_c \Delta E}{2}} \, ,
	\end{equation}
with $0<\omega<1$.
For later convenience, we also define
	\begin{equation}
	\label{eq:Epsilon}
	\epsilon = \frac{\omega - \sigma}{1 - \omega \sigma} = \tanh{\frac{(\beta_c - \beta_h) \Delta E}{ 2}}  \, ,
	\end{equation} 
whose value, $0<\epsilon<1$, quantifies the temperature difference between the two reservoirs.

Finally, we assume that the incoming bit stream contains a mixture of 0's and 1's, with probabilities $p_0$ and $p_1$, respectively, with no correlations between bits.
Let
\begin{equation}
\label{eq:Delta}
\delta \equiv p_0-p_1
\end{equation}
denote the proportional excess of 0's among incoming bits.

We thus have the following dynamics.
When a fresh bit arrives to interact with the demon, its state is $0$ or $1$.
The demon and bit subsequently interact for a time $\tau$, making the transitions shown in Fig.~\ref{fig:Setup}(b), thereby exchanging energy with the reservoirs.
The state of the bit at the end of the interaction interval is then preserved as the bit joins the outgoing stream, and the next bit in the sequence moves in to 
have its turn with the demon.
The parameters $\gamma$, $\sigma$ and $\omega$ define the intrinsic and cooperative transition rates (Eqs.~\ref{eq:sigma_gamma}, \ref{eq:omega}), $\tau$ gives the duration of interaction with each bit, and $\delta$ specifies the statistics of the incoming bits.
Under these dynamics, the demon evolves to a periodic steady state, in which its behavior is statistically the same from one interaction interval to the next.

Before proceeding to the solution of these dynamics, we discuss heuristically how our model can achieve the systematic transfer of heat from the cold to the hot reservoir.
For this purpose let us assume that each incoming bit is in state $0$, hence $\delta=1$.
At the start of a particular interaction interval, the joint state of the demon and newly arrived bit is either $0u$ or $0d$.
The demon and bit then evolve together for a time $\tau$, according to the transitions shown in Fig.~\ref{fig:Setup}(b).
If the joint state at the end of the interaction interval is $0u$ or $0d$, then it must be the case that every transition $0d \rightarrow 1u$ was balanced by a transition $0d \leftarrow 1u$, hence no net energy was absorbed from the cold reservoir.
If the final state is $1u$ or $1d$, then we can infer that there was one {\it net} transition from $0d$ to $1u$, and a quantity of energy $\Delta E$ was absorbed from the cold reservoir.
This amounts to {\it thermal rectification}: over the course of one interaction interval, energy can be withdrawn from the cold reservoir but not delivered to it.
Moreover, a record of this process is imprinted in the bit stream, as every outgoing bit in state 1 indicates the absorption of energy $\Delta E$ from the cold reservoir.
Since the demon also exchanges energy with the hot reservoir, and since energy cannot accumulate indefinitely within the demon, in the long run we get a net flux of energy from the cold to the hot reservoir, proportional to the rate at which 1's appear in the outgoing bit stream.

More generally, if the incoming bit stream contains a mixture of 0's and 1's, then an excess of 0's (that is, $\delta > 0$) produces a statistical bias that favors the flow of heat from the cold to the hot reservoir, while an excess of 1's ($\delta < 0$) produces the opposite bias.
This bias either competes with or enhances the normal thermodynamic bias due to the temperature difference between the two reservoirs.
The demon thus affects the flow of energy between the reservoirs, and modifies the states of the bits in the memory register.
We now investigate quantitatively the interplay between these two effects.

Once the demon has reached its periodic steady state, let $p_0^\prime$ and $p_1^\prime$ denote the fractions of 0's and 1's in the outgoing bit stream, and let $\delta^\prime = p_0^\prime - p_1^\prime$ denote the excess of outgoing 0's.
Then
	\begin{equation}
	\label{eq:Phi}
	\Phi \equiv p_1' - p_1 = \frac{\delta - \delta'}{2}
	\end{equation}
represents the average production of 1's per interaction interval in the outgoing bit stream, relative to the incoming bit stream.
Since each transition $0\rightarrow 1$ is accompanied by the absorption of energy $\Delta E$ from the cold reservoir (Fig.~\ref{fig:Setup}(b)), the average transfer of energy from the cold to the hot reservoir, per interaction interval, is given by
	\begin{equation}
	\label{eq:QPhi}
	Q_{c \rightarrow h} =  \Phi \Delta E \, .
	\end{equation}
A positive value of $Q_{c \rightarrow h}$ indicates that our device pumps energy against a thermal gradient, like the creature imagined by Maxwell.

To quantify the information-processing capability of the demon, let
	\begin{equation}
	\label{eq:S}
	S(\delta) = -\sum_{i=0}^{1} p_i \ln p_i = 
	- \frac{1 - \delta}{2} \ln{\frac{1 - \delta}{2}} - \frac{1 + \delta}{2} \ln{\frac{1 + \delta}{2}} \, 
	\end{equation} 
denote the information content, per bit, of the incoming bit stream, and define $S(\delta^\prime)$ by the same equation, for the outgoing bit stream.
Then
	\begin{equation}
	\label{eq:DeltaS_B}
	\Delta S_B \equiv S(\delta') - S(\delta) = S(\delta - 2 \Phi) - S(\delta)
	\end{equation}
provides a measure of the extent to which the demon increases the information content of the memory register.
We will interpret a positive value of $\Delta S_B$ to indicate that the demon {\it writes} information to the bit stream, while a negative value indicates {\it erasure}.
(More precisely, since $S(\delta^\prime)$ neglects the small correlations that arise between the outgoing bits, $\Delta S_B$ reflects the change in the Shannon information of the {\it marginal} probability distribution of each outgoing bit.)

From Eqs.~\ref{eq:QPhi} and \ref{eq:DeltaS_B} we see that $\Phi$ determines both $Q_{c \rightarrow h}$ and $\Delta S_B$.
In the Supplemental Material, we show that under the dynamics we have described, the demon reaches a periodic steady state, determined by the model parameters $\Lambda \equiv (\delta, \sigma, \gamma, \omega, \tau)$, in which
	\begin{equation}
	\label{eq:Phi_solution}
	 \Phi(\Lambda) = \frac{\delta - \epsilon}{2} \,\, \eta(\Lambda) \quad , \quad \eta > 0
	 \end{equation}
and
	\begin{equation}
	\label{eq:SecondLaw}
	Q_{c \rightarrow h} ( \beta_h - \beta_c) + \Delta S_B \geq 0 \, .
	\end{equation}
Eq.~\ref{eq:SecondLaw} is a strict inequality when $\delta\ne\epsilon$.
An explicit expression for $\eta(\Lambda)$ is given in the Supplemental Material, but for our present purposes the crucial point is that the {\it sign of $\Phi$ is the same as that of} $\delta - \epsilon$.
We can think of two effective forces: the bias induced by the incoming bit stream, which favors $\Phi > 0$ when $\delta > 0$ (as discussed above), and the temperature gradient, quantified by $\epsilon$, which favors $\Phi < 0$ (Eq.~\ref{eq:QPhi}).
When these compete, the winner is determined by the difference $\delta-\epsilon$.

Eq.~\ref{eq:Phi_solution} is obtained by solving for the periodic steady state of the demon, using a linear-algebraic approach.
Eq.~\ref{eq:SecondLaw} is obtained by constructing a Lyapunov function for the demon and interacting bit.
The details of these derivations are provided in the Supplemental Material.
Here, we instead use these results to investigate the behavior of our model in the periodic steady state.
To that end, we fix $\gamma$ and $\omega$ and construct a phase diagram that illustrates the dependence on $\delta$ and $\epsilon$, for various values of $\tau$, shown in Fig.~\ref{fig:Phase}.
Let us consider the different regions of this diagram, working our way from right to left.

From Eqs.~\ref{eq:QPhi} and \ref{eq:Phi_solution} it follows that $Q_{c \rightarrow h} > 0$ when $\delta>\epsilon$, shown as the most darkly shaded region in Fig.~\ref{fig:Phase}.
Here, a surplus of incoming 0's prevails over the temperature difference and our demon generates a flow of energy from the cold to the hot reservoir.
Moreover, Eq.~\ref{eq:SecondLaw} reveals that $\Delta S_B>0$ in this region (since $\beta_h<\beta_c$).
This agrees with the consensus described earlier: in order for a physical device to act in the manner of Maxwell's demon, it must write information to a physical memory register.
In this sense, a bit stream with a low information content can be viewed as a thermodynamic resource, which can be expended (by writing to the available memory) in order to achieve refrigeration.

Now consider the region $\epsilon>\delta>0$, in which the surplus of 0's in the incoming bit stream is not sufficient to overcome the temperature gradient, and energy flows from the hot to the cold reservoir.
Since $\Phi<0$ we get $\delta^\prime>\delta>0$ (Eq.~\ref{eq:Phi}).
This in turn implies $\Delta S_B<0$, as $S(\delta)$ is a concave function with a maximum at $\delta=0$.
In this region the demon acts as an eraser, lowering the information content of the bit stream, but the price paid for this erasure is the passage of heat from the hot to the cold reservoir.

In the region $\delta<0$, energy flows from the hot to the cold reservoir (Eqs.~\ref{eq:QPhi}, \ref{eq:Phi_solution}), but the value of $\Delta S_B$ depends on all the model parameters.
In Fig.~\ref{fig:Phase}, for four different values of $\tau$, we show the line corresponding to $\Delta S_B=0$.
To the right of this line we have $\Delta S_B <0$ and to the left we have $\Delta S_B>0$.
In the limit $\tau\rightarrow\infty$, the boundary between these two behaviors approaches the line $\epsilon = -\delta$.

Examining the phase diagram as a whole, we see that in the shaded regions our model reaches a steady state in which one thermodynamic resource is replenished at the expense of another.
Either energy is pumped against a thermal gradient at the cost of writing information to memory (the refrigerator regime),
or else memory is made available, by erasure, at the expense of allowing energy to flow from the hot to the cold reservoir (the eraser regime).
The boundary between these two behaviors is the line $\delta=\epsilon$.
In the unshaded region at the far left, both resources are consumed, as energy flows down the thermal gradient and information is written to the bit stream.

Finally, to place our model within the context of the second law of thermodynamics, note that the first term on the left side of Eq.~\ref{eq:SecondLaw} is the steady-state change in thermodynamic entropy due to the flow of heat, and the second term is the change in information entropy, per interaction interval.
Eq.~\ref{eq:SecondLaw} can be viewed as a modified Clausius inequality, in which the information entropy of a random sequence of data is explicitly assigned the same thermodynamic status as the physical entropy associated with the transfer of heat.
(More precisely, Eq.~\ref{eq:SecondLaw} is a weak version of this inequality, as we neglect correlations among the outgoing bits; see Supplemental Material.)
Thus our model provides support for the consensus mentioned earlier~\cite{Landauer1961,Penrose1970,Bennett1982}, and Eq.~\ref{eq:SecondLaw} is consistent with Landauer's principle~\cite{Landauer1961}, which states that a thermodynamic cost must be paid for the erasure of memory.
However, in Ref.~\cite{Landauer1961} this cost appears as the dissipation of energy into a single thermal reservoir, whereas in our model it is the transfer of energy from a hot to a cold reservoir.

In summary, we have constructed a simple, solvable model of an autonomous physical system that mimics the behavior of the ``neat-fingered being'' in Maxwell's thought experiment, generating a systematic flow of energy against a thermal gradient without the input of external work.
While Maxwell's creature accomplishes this with intelligence, our inanimate device requires only a memory register to which information can be written.
Alternatively, it can harness the flow of energy from hot to cold in order to erase information from the register.

	   We thank Andy Ballard, Shaon Chakrabarti, Sebastian Deffner, and Zhiyue Lu for useful discussions, and gratefully acknowledge financial support from the National Science Foundation (USA) under grants DMR-0906601, ECCS-0925365, and DMR-1206971, the University of Maryland, College Park, and Peking University.

\begin{widetext}

\section{Supplemental Material}

\appendix

\subsection{Solving for $\Phi(\Lambda)$}
\label{app:derivation}

Solving for $\Phi$ involves first solving for the periodic steady state of the demon, then using that solution to determine the distribution of the outgoing bits,  from which $\Phi$ follows by Eq.~6 of the main text.
We will use the notation ${\bf p}^D = (p_u,p_d)^T$ (where the superscript $T$ indicates transpose) to denote the statistical state of the demon, ${\bf p}^B = (p_0,p_1)^T$ for that of the interacting bit, and ${\bf p} = (p_{u0},p_{d0},p_{u1},p_{d1})^T$ to denote their joint probability distribution.

Let $\mathcal{T}$ denote the $2 \times 2$ transition matrix whose element $T_{\mu \nu}$ ($\mu, \nu \in \{u, d\}$) gives the probability for the demon to be in state $\mu$ at the end of an interaction interval, given that it was in state $\nu$ at the start of the interval.
As explained below, this matrix can be written as the product
\begin{equation}
\label{eq:T}
\mathcal{T} = \mathcal{P}^D e^{\mathcal{R} \tau} \mathcal{M} ,
\end{equation}
where
\begin{equation}
\label{eq:Matrices}
\mathcal{P}^D =
\left(
\begin{array}{cccc}
1 & 0 & 1 & 0 \\
0 & 1 & 0 & 1
\end{array}
\right)
, \quad
\mathcal{R} = 
\left(
\begin{array}{cccc}
\bullet & \gamma (1-\sigma ) & 0 & 0 \\
\gamma (1+\sigma ) & \bullet & 1+\omega  & 0 \\
0 & 1-\omega  & \bullet & \gamma (1-\sigma ) \\
0 & 0 & \gamma (1+\sigma ) & \bullet 
\end{array}
\right)
, \quad
\mathcal{M} = 	
\left(
\begin{array}{cc}
p_0 & 0 \\
0 & p_0 \\
p_1 & 0 \\
0 & p_1
\end{array}
\right) .
\end{equation}
Here ${\cal R}$ is the transition rate matrix for the demon and the interacting bit.
Its off-diagonal elements are given by Eqs.~2 and 3 of the main text, and its diagonal elements are determined by the requirement that the elements in each column sum to zero~\cite{vanKampen2007}.
To understand Eq.~\ref{eq:T}, let $\mathbf{p}_0^{D}$ denote the distribution of the demon at the start of a given interaction interval.
Then ${\bf p}_0 = \mathcal{M} {\bf p}_0^{D}$ gives the initial joint distribution of the demon and the incoming bit.
From this initial, uncorrelated state the joint distribution evolves under the master equation ${\mathrm d}{\bf p}/{\mathrm d}t = \mathcal{R} {\bf p}$, therefore ${\bf p}_\tau=\exp{(\mathcal{R} \tau)} \mathcal{M} \mathbf{p}_0^{D}$ gives the joint distribution at the end of the interaction interval.
The matrix $\mathcal{P}^D$ then projects out the state of the bit, thus ${\bf p}_\tau^D = \mathcal{P}^D \exp{(\mathcal{R} \tau)} \mathcal{M} \mathbf{p}_0^{D} = \mathcal{T} \mathbf{p}_0^{D}$ gives the final marginal distribution of the demon.

The evolution of the demon over many intervals is given by repeated application of  the matrix $\mathcal{T}$.
Because $\mathcal{T}$ is a positive transition matrix~\cite{Meyer2000}, the demon evolves to a periodic steady state,
\begin{equation}
\lim_{n\rightarrow\infty} \mathcal{T}^n \mathbf{p}_0^{D} = \mathbf{p}_0^{D, ps} \quad ,
\end{equation}
defined uniquely by
\begin{equation}
\label{eq:eigenstate_of_T}
\mathcal{T} \mathbf{p}_0^{D, ps} = \mathbf{p}_0^{D, ps} \quad.
\end{equation}
$\mathbf{p}_0^{D, ps}$ gives the marginal distribution of the demon at the start of each interaction interval.

In the periodic steady state, the joint distribution of the demon and the interacting bit, at the end of the interaction interval, is given by ${\bf p}_\tau^{ps} = \exp{(\mathcal{R} \tau)} \mathcal{M} \mathbf{p}_0^{D,ps}$.
The marginal distribution of the outgoing bit is then given by projecting out the state of the demon:
\begin{equation}
\label{eq:pBf}
\mathbf{p}_\tau^{B, ps} = \mathcal{P}^B e^{\mathcal{R} \tau} \mathcal{M} \mathbf{p}_0^{D, ps} 
\quad, \quad
\mathcal{P}^B \equiv
\left(
\begin{array}{cccc}
1 & 1 & 0 & 0\\
0 & 0 & 1 & 1
\end{array}
\right).
\end{equation}

Therefore, to solve for $\Phi$, we first compute the elements of $\mathcal{T}$ using Eq.~\ref{eq:T}, then find its right eigenstate $\mathbf{p}_0^{D,ps}$ (Eq.~\ref{eq:eigenstate_of_T}), then determine $\mathbf{p}_\tau^{B, ps} = (p_0^\prime,p_1^\prime)^T$ using Eq.~\ref{eq:pBf}.
$\Phi$ then follows directly from Eq.~6 in the main text: $\Phi = p_1' - p_1$.

We performed these calculations using Mathematica~\cite{Mathematica8}, and then simplified the results substantially by hand, finally obtaining
\begin{subequations}
\label{eq:PhiAll}
\begin{equation}
\label{eq:PhiGen}
\Phi = \frac{\delta - \epsilon}{2} \, \eta(\Lambda) \quad, \quad \eta(\Lambda) = \frac{\nu_2 P + \nu_3 Q}{P + Q},
\end{equation}
\begin{equation}
\label{eq:defs}
\begin{array}{ll}
P = \mu_2\,(\mu_4 \nu_3 + \mu_1 \nu_1) \quad &, \quad Q = \mu_3 \, (\mu_4 \nu_2 + \mu_1 \nu_1),\\
\nu_1 = 1 - e^{-2 \gamma \, \tau }\quad  & ,  \quad \mu_1 = (\delta +\sigma ) \, \omega , \\
\nu_2 = 1 - e^{-(1 + \gamma  - \alpha ) \, \tau } \quad  & ,  \quad \mu_2  = \alpha  + \gamma  + \sigma \, \omega, \\
\nu_3 = 1 - e^{-(1 + \gamma + \alpha ) \, \tau } \quad & , \quad \mu_3  = \alpha  - \gamma  - \sigma \, \omega , \\
\alpha = \sqrt{1+\gamma ^2+2 \gamma  \sigma  \omega }\quad & , \quad \mu_4  = 1- \delta \, \omega.
\end{array}
\end{equation}
\end{subequations}

If the demon's intrinsic transitions occur rapidly in comparison with the cooperative transitions, $\gamma\rightarrow\infty$, then the analysis simplifies substantially: the demon remains in equilibrium with the hot reservoir at all times, and the interacting bit obeys the master equation
	\begin{equation}
	\label{eq:GammaInfty}
	\frac{\rm d}{{\rm d}t}
	\left(
	\begin{array}{c}
	{p}_0^B\\
	{p}_1^B
	\end{array}
	\right)
	 = 
	 \left(
	 \begin{array}{cc}
	 -a & \,\,\,\,b \\
	 \,\,\,a & -b
	 \end{array}
	 \right)
	 \left(
	 \begin{array}{c}
	 p_0^B \\
	 p_1^B
	 \end{array}
	 \right) \, ,
	\end{equation}
with $a=(1-\omega)(1+\sigma)/2$ and $b=(1+\omega)(1-\sigma)/2$.
Here $p_j^B(t)$ is the probability to find the bit in state $j\in\{0,1\}$ at time $t$ during the interaction interval.
Integrating Eq.~\ref{eq:GammaInfty} over one interaction interval, $0 \le t \le \tau$, then setting $p_1 = p_1^B(0)$ and $p_1^\prime = p_1^B(\tau)$ in Eq.\ 6 of the main text, we obtain
\begin{equation}
\Phi = \frac{\delta - \epsilon}{2} \, \left[ 1 - e^{- (1 - \sigma \omega) \tau} \right] \, .
\end{equation}
As a consistency check, we note that this result also follows from our general solution, Eq.~\ref{eq:PhiAll}, with the expression for $\eta(\Lambda)$ evaluated in the limit $\gamma\rightarrow\infty$.

Our general expression for $\eta(\delta, \sigma, \gamma, \omega, \tau)$, while exact, is sufficiently complex that we are unable to derive the inequality $\eta>0$ (which was crucial in our interpretation of the phase diagram in the main text) directly from Eq.~\ref{eq:PhiAll}.
Instead we will show in Appendix~\ref{sec:eta} that this inequality follows from the modified Clausius inequality, Eq.~11 of the main text, which we now derive.

\subsection{Modified Clausius inequality}
\label{app:Clausius}

During any interaction interval, the joint distribution of the demon and the interacting bit evolves according to the master equation discussed above,
\begin{equation}
\label{eq:master}
\frac{{\mathrm d}{\bf p}}{{\mathrm d}t} =\mathcal{R} {\bf p},
\end{equation}
where $\mathcal{R}$ is given in Eq.~\ref{eq:Matrices}.
For very long interaction intervals ($\tau\rightarrow\infty$), the combined system relaxes to the stationary state
\begin{equation}
\label{eq:ss_expression}
\overline{\bf p} = \frac{1}{\mathcal{N}} \left( 1,  \mu , \mu \nu, \mu^2 \nu \right)^T, \quad \mu = \frac{1 + \sigma}{1 - \sigma}, \quad \nu = \frac{1 - \omega}{1 + \omega},
\quad
\mathcal{N} = (1 + \mu)(1 + \mu \nu) \, ,
\end{equation}
which satisfies $\mathcal{R}\overline{\bf p} = {\bf 0}$.
Note that $\overline{\bf p}$ is actually a product of marginal distributions $\overline{\bf p}^{D}$ and $\overline{\bf p}^{B}$ for the demon and bit:
\begin{subequations}
\label{eq:product}
\begin{equation}
\label{eq:product_a}
\overline p_{ij} = \overline p^{D}_i \, \overline p^{B}_j,  \quad i \in \{u, d\}, \quad j \in \{0, 1\},
\end{equation}
\begin{equation}
\label{eq:product_b}
\overline {\bf p}^{D} = (1, \mu)^T / (1 + \mu) , \quad \overline {\bf p}^{B} = (1, \mu \nu)^T / (1 + \mu \nu).
\end{equation}
\end{subequations}
The irreversible approach of ${\bf p}(t)$ toward $\overline{\bf p}$ is described by the \emph{relative entropy}~\cite{Cover2006},
\begin{equation}
\label{eq:D}
D({\bf p} || \overline{\bf p}) = \sum_k p_k \ln{\frac{p_k}{\overline p_k}} \ge 0 \, .
\end{equation}
Here and in what follows, we use the index $k$ to indicate a joint state of the demon and the bit, $k \in \{0u, 0d, 1u, 1d\}$, reserving $i$ and $j$ for the demon and the bit, respectively, as in Eq.~\ref{eq:product_a}.
A standard calculation~\cite{vanKampen2007} shows that $D$ is a Lyapunov function, that is it satisfies
\begin{equation}
\label{eq:monotonicRelaxation}
\frac{\mathrm{d}}{\mathrm{d}t} \, D({\bf p} || \overline{\bf p}) \leq 0 \, ,
\end{equation}
where the equality holds only when ${\bf p} = \overline{\bf p}$.
Thus, as measured by relative entropy, any initial ${\bf p} \neq \overline{\bf p}$ evolves monotonically toward $\overline{\bf p}$, although for finite interaction intervals this relaxation is interrupted by the arrival of the next bit.
We now use these properties to derive the inequality
\begin{equation}
\label{eq:Clausius}
Q_{c \rightarrow h} (\beta_h - \beta_c) + \Delta S_B  \geq  0 \, ,
\end{equation}
which appears as Eq.~11 of the main text.

Let ${\bf p}_0$ and ${\bf p}_\tau$ denote the joint distributions of the demon and a bit at the beginning and end of a given interaction interval, respectively, and similarly define ${\bf p}_0^D$, ${\bf p}_\tau^D$, ${\bf p}_0^B$ and ${\bf p}_\tau^B$ for the marginal distributions of the demon and the bit.
Eq.~\ref{eq:monotonicRelaxation} implies 
\begin{equation}
\label{eq:monotonic}
 D({\bf p}_0 || \overline{\bf p}) - D({\bf p}_\tau || \overline{\bf p}) \geq 0.
\end{equation}
Using Eqs.~\ref{eq:D} and \ref{eq:product_a} we rewrite the left side of this equation as 
\begin{equation}
\label{eq:expansion}
S_\tau - S_0 - \sum_{i \in \{u, d\}} \left( p_{\tau,i}^{D} - p_{0,i}^{D} \right) \ln{\overline p^{D}_i} - \sum_{j \in \{0, 1\}} \left( p_{\tau,j}^{B} - p_{0,j}^{B} \right) \ln{\overline p^{B}_j},
\end{equation}
where $S_0 = -\sum_k p_{0,k} \ln p_{0,k}$ and $S_\tau = -\sum_k p_{\tau,k} \ln p_{\tau,k}$ are the information entropies of the joint distributions of the demon and the bit at the beginning and end of the interaction interval.
Let us now evaluate Eq.~\ref{eq:expansion}, assuming the demon has reached its periodic steady state.

The joint entropy $S$ can be written as~\cite{Cover2006}
\begin{equation}
\label{eq:decomposition}
S = S^D + S^B - I(D;B) \quad, \quad I(D;B) \geq 0,
\end{equation}
where $S^D$ is the marginal entropy of the demon, $S^B$ is the marginal entropy of the bit, and the {\it mutual information} $I(D;B)$ quantifies the degree of correlation between them.
By construction, the demon and bit are uncorrelated at the start of the interaction interval, hence $I_0(D;B) = 0$.
In the periodic steady state we have $S_\tau^{D} = S_0^{D}$, because the demon starts and ends in the same distribution.
Hence the difference $S_\tau - S_0$ in Eq.~\ref{eq:expansion} can be replaced by $\Delta S_B - I_\tau(D;B)$.
We also have ${\bf p}_0^{D}={\bf p}_\tau^{D}$ in the periodic steady state, so the first sum appearing in Eq.~\ref{eq:expansion} vanishes. 

Once the period steady state has been reached, the bit distributions ${\bf p}_0^B$ and ${\bf p}_\tau^B$ correspond to the statistics of the incoming and outgoing bit streams, defined in the main text:
\begin{equation}
p_{0,j}^B = p_j \quad,\quad p_{\tau,j}^B = p_j^\prime \quad,\quad j \in \{0,1\} \, ,
\end{equation}
hence $p_{\tau,0}^{B} - p_{0,0}^{B} = - (p_{\tau,1}^{B} - p_{0,1}^{B}) = \Phi$, from the definition of $\Phi$.
The last term in Eq.~\ref{eq:expansion} can now be rewritten, using Eq.~\ref{eq:ss_expression} and Eqs.~2 and 3 of the main text, as
\begin{equation}
\label{eq:munu}
-\sum_{j \in \{0, 1\}} \left( p_{\tau,j}^{B} - p_{0,j}^{B} \right) \ln{\overline p^{B}_j} = \Phi \ln(\mu \nu) = Q_{c \rightarrow h} (\beta_h - \beta_c).
\end{equation}
Collecting these results, we get
\begin{equation}
\label{eq:expansion_2}
D({\bf p}_0 || \overline{\bf p}) - D({\bf p}_\tau || \overline{\bf p}) = \Delta S_B - I_\tau(D;B) + Q_{c \rightarrow h} (\beta_h - \beta_c),
\end{equation}
which then combines with Eq.~\ref{eq:monotonic} to give us
\begin{equation}
\label{eq:protoClausius}
Q_{c \rightarrow h} (\beta_h - \beta_c) + \Delta S_B  \geq I_\tau(D;B) \geq 0.
\end{equation}
An alternative derivation of this result can be constructed using the integral fluctuation theorem for total entropy production~\cite{Seifert2005}.

The first inequality in Eq.~\ref{eq:protoClausius} is stronger than the modified Clausius statement, Eq.~\ref{eq:Clausius}.
This underscores the fact that Eq.~\ref{eq:Clausius} is a weak statement of the second law of thermodynamics (as it applies to our model), since it neglects correlations in the outgoing bits: the quantity $\Delta S_B$ is defined in terms of the marginal distribution of each bit.
In reality the bits do develop correlations via their interactions with the demon, as the state of the demon at the end of one interaction interval is also its initial state at the beginning of the next interval.
(Explicit numerical simulations indicate that these correlations are small, but not zero.)
If these correlations were to be taken into account, then the net change in the Shannon entropy per bit would have a value slightly lower than $\Delta S_B$, and Eq.~\ref{eq:Clausius} would be replaced by a somewhat stronger bound.
These considerations are reflected, somewhat indirectly, by the term $I_\tau(D;B)$ in Eq.~\ref{eq:protoClausius}.

Finally, note that
\begin{equation}
\frac{\overline p_1^B}{\overline p_0^B} = \mu\nu = \frac{1-\epsilon}{1+\epsilon}
\quad,\quad
\frac{p_1}{p_0} = \frac{1-\delta}{1+\delta} \, ,
\end{equation}
using Eqs.~\ref{eq:ss_expression} and \ref{eq:product_b}, and the definitions of $\epsilon$ and $\sigma$.
Thus, when $\delta=\epsilon$, the incoming bits arrive in the stationary distribution $\overline{\bf p}$.
In this case, no relaxation occurs during the interaction interval; the equality holds in Eqs.~\ref{eq:monotonicRelaxation} and \ref{eq:monotonic}; the outgoing bits depart with the same distribution; and $\Phi=0$.
When $\delta\ne\epsilon$, Eqs.~\ref{eq:monotonicRelaxation} and \ref{eq:monotonic} are both strict inequalities, and therefore so is the modified Clausius inequality (Eq.~\ref{eq:Clausius} / Eq.~11).

\section{Positivity of $\eta(\Lambda)$}
\label{sec:eta}

\begin{figure}
\subfigure[]{
\includegraphics[scale = 0.35]{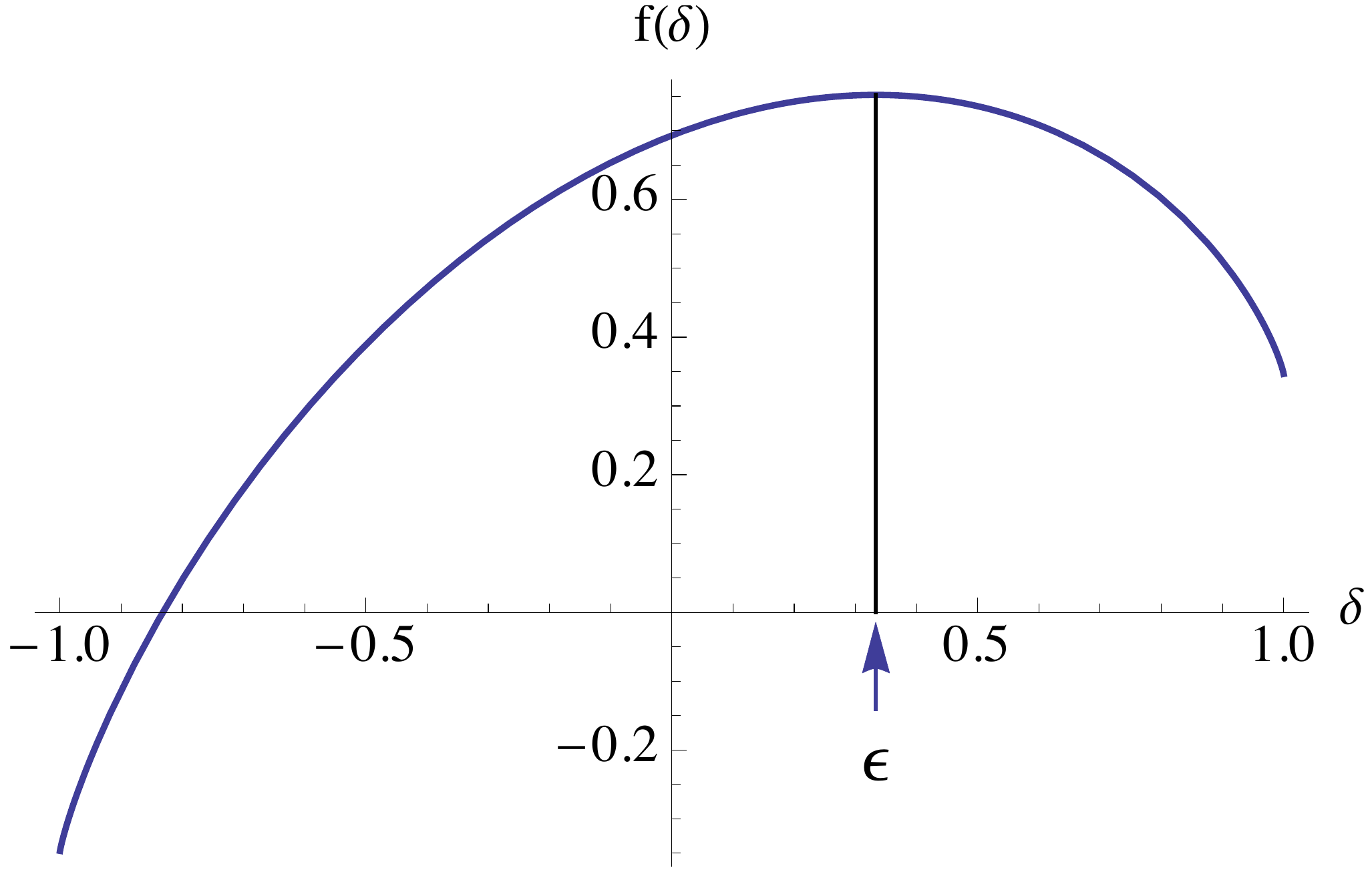}}
\subfigure[]{
\includegraphics[scale = 0.35]{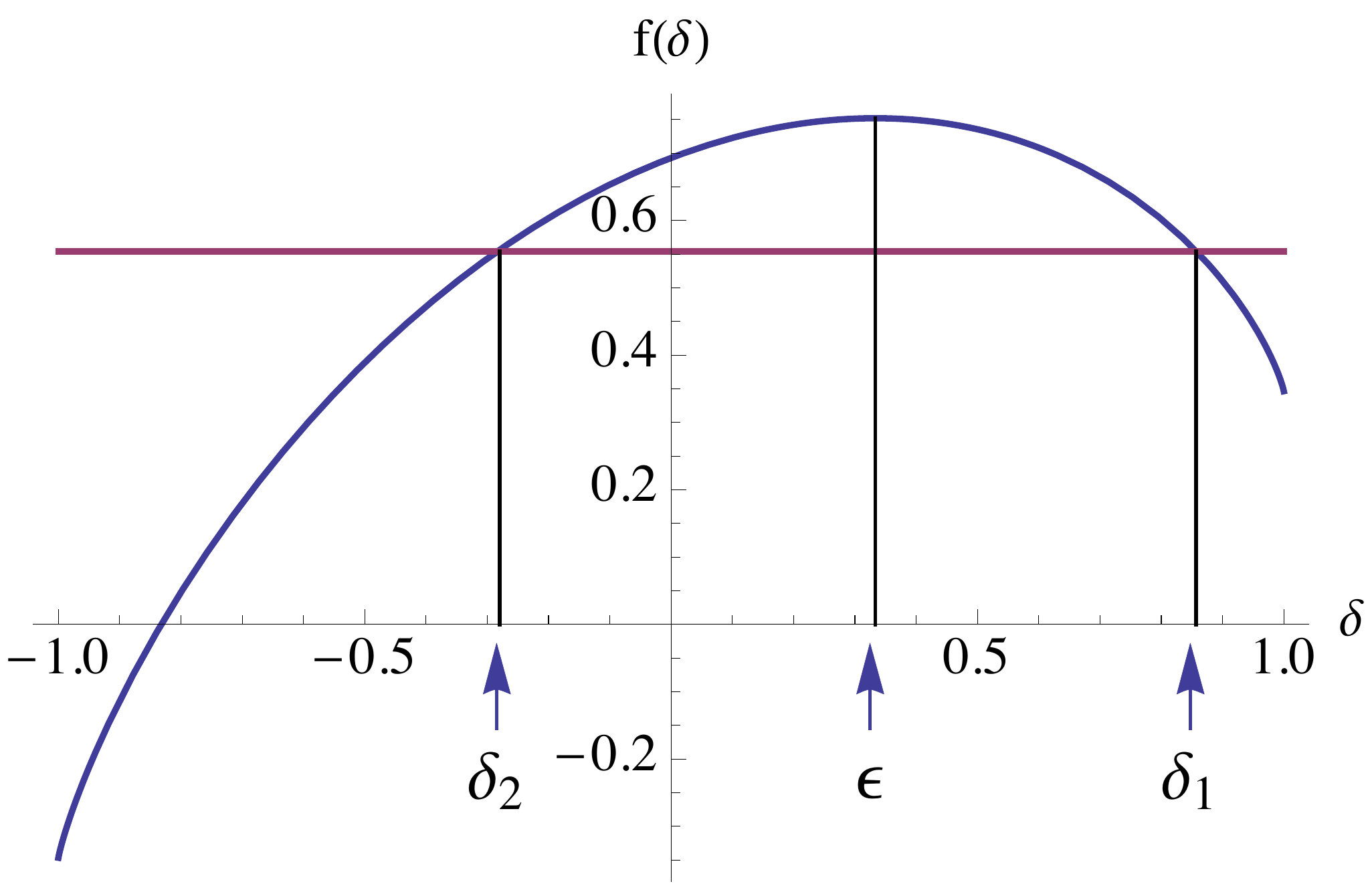}}
\caption{(a) The concave function $f(\delta)$ has a maximum at $\delta = \epsilon$, as illustrated for $\epsilon = 1/3$. (b) For a given $\delta_1$, we must have $\delta_2 < \delta_1^\prime < \delta_1$ to ensure $f(\delta'_1) > f(\delta_1)$. Hence, both $\delta_1^\prime$ and $\epsilon$ lie to the left of $\delta_1$.}
\label{fig:f}
\end{figure}

To investigate the sign of $\eta$, let us take $\delta\ne\epsilon$~\footnote{
When $\delta=\epsilon$, the value of $\eta$ is inconsequential, by Eq.~\ref{eq:PhiGen}.} and rewrite Eq.~\ref{eq:Clausius} in the form
\begin{equation}
\label{eq:f}
f(\delta') > f(\delta) \, ,
\end{equation}
where
\begin{equation}
\label{eq:f_defs}
f(\delta) = K \delta + S(\delta) \quad , \quad K = \frac{1}{2} (\beta_c - \beta_h) \Delta E > 0.
\end{equation}
Eq.~\ref{eq:f} follows by the direct substitution of the relations
\begin{equation}
\label{eq:C_defs}
Q_{c \rightarrow h} = \Phi \Delta E \quad , \quad \Phi = \frac{\delta - \delta'}{2} \quad , \quad \Delta S_B = S(\delta') - S(\delta)
\end{equation}
into Eq.~\ref{eq:Clausius}, using a strict inequality since $\delta\ne\epsilon$.

By construction, ${\rm d}^2f/{\rm d}\delta^2 < 0$.
Setting ${\rm d}f/{\rm d}\delta=0$, the unique maximum of $f(\delta)$ is easily shown to occur at $\delta = \epsilon$, as illustrated in Fig.~\ref{fig:f}(a) for $\epsilon = 1/3$.
Now let $\delta_1$ and $\delta_2$ denote two values of $\delta$ that correspond to the same value of $f$, with $\delta_2 < \epsilon < \delta_1$, as shown in Fig.~\ref{fig:f}(b).
Let $\delta_1^\prime$ describe the surplus of $0$'s in the outgoing bit stream, when the incoming stream is characterized by $\delta_1$.
Because the maximum of $f(\delta)$ occurs at $\delta=\epsilon$, Eq.~\ref{eq:f} implies that $\delta_2 < \delta_1^\prime < \delta_1$; see Fig.~\ref{fig:f}(b).
If we instead consider incoming and outgoing bit streams described by $\delta_2$ and $\delta_2^\prime$, then the same argument gives us $\delta_2 < \delta_2^\prime < \delta_1$.
We therefore conclude that the incoming and outgoing bit streams necessarily satisfy
\begin{equation}
\label{eq:sign}
\text{sign}(\delta - \delta^\prime) = \text{sign}(\delta - \epsilon) \, ,
\end{equation}
in other words $\delta^\prime$ lies on the same side as $\epsilon$ with respect to $\delta$.
Since 
\begin{equation}
\label{eq:Phi_def}
\frac{\delta-\delta^\prime}{2} = \Phi = \frac{\delta - \epsilon}{2} \eta(\Lambda) \, ,
\end{equation}
we must have $\eta(\Lambda)>0$.

\end{widetext}

	\end{document}